\documentclass[a4paper,11pt]{article}
\usepackage{epsfig}
\usepackage{amssymb}
\usepackage{amsmath}

\begin{document}

\title{Study of Temperature Inversion Symmetry for the Twisted Wess-Zumino}
\author{V.K.Oikonomou\thanks{%
voiko@physics.auth.gr}\\
%EndAName
Dept. of Theoretical Physics Aristotle University of Thessaloniki,\\
Thessaloniki 541 24 Greece}
\maketitle

\begin{abstract}
The temperature inversion symmetry, for a non interacting
supersymmetric ensemble, at finite volume, is studied. It is found
that, the scaled free energy, $f(\xi )$, is antisymmetric under
temperature inversion transformation, {\it{i.e.}}~$f(\xi )=-\xi
^{d}f(\frac{1}{\xi })$. This occurs for antiperiodic bosons and
periodic fermions, in the compact dimension. On the contrary, for
periodic bosons and antiperiodic fermions, $f(\xi )=\xi
^{d}f(\frac{1}{\xi })$.
\end{abstract}

\bigskip
\bigskip
\section*{Introduction}

One of the most interesting phenomena in Quantum Field Theory is
the Casimir effect. It expresses the quantum fluctuations of the
vacuum of a quantum field. It originates from the "confinement" of
a field in finite volume. Many studies have been done since H.
Casimir's original work. The Casimir energy, usually calculated in
these studies, is closely related to the boundary conditions of
the fields under consideration. Boundary conditions influence the
nature of the so-called Casimir force, which is generated from the
vacuum energy.

By calculating the Casimir energy at finite temperature, one finds
many interesting properties of fermionic and bosonic fields. In
ref.~\cite{wotzasek} the temperature inversion symmetry of the
Casimir energy, for a spin $0$ bosonic field, at finite
temperature, was studied. The author actually examined some
thermodynamic quantities for a bosonic field, at finite
temperature and with a compact
space dimension, in three spatial dimensions, {\it{i.e.}}~a space having $%
S^{1}\times R^{2}$ spatial topology. He found that the quantity,
\begin{equation}
f(\xi )=-\frac{L^{d-1}}{\Gamma (\frac{d}{2})\xi ^{d}\pi ^{-\frac{d}{2}}}F,
\label{wotz1}
\end{equation}%
{\it{i.e.}}~the scaled free energy $F$ of a spin $0$ bosonic field, with $F$,%
\begin{equation}
F\sim T\sum_{n,m} \int dk^{2}\ln [k^{2}+(2\pi nT)^{2}+(\frac{2\pi m}{L}%
)^{2}],  \label{wotz}
\end{equation}%
is invariant under the transformation $L\rightarrow \frac{1}{T}$
($L$ is the length of the compact dimension). As it can be easily
seen from relation (\ref{wotz}), he used periodic boundary
conditions, corresponding to the compact dimension and periodic
for the "temperature dimension" (consistent with the KMS
relations). In previous and
later results \cite{Santos,Santos1,Pinto,Tollefsen,Gundersen,Brown}, the symmetries that the function (\ref%
{wotz1}) has, were studied and calculations for other fields, such
as fermions and gauge fields, with various boundary conditions,
were done. Studying thermodynamical quantities of fields in such
topological spaces is of great importance (for
example in microelectronics where characteristic distances become small \cite%
{Srivastava}).

In this work, the free energy for an ensemble of massless, non
interacting
\ bosons and fermions, with equal degrees of freedom, in spaces with $\frac{%
R^{1}}{Z_{\infty }}\times R^{d-2}$ spatial topologies at finite
temperature, will be examined ($d$ is the total dimension of the
space-time before compactification, $Z_{\infty }$ the infinite
cyclic group). Specifically, the temperature inversion symmetry
shall be studied for this ensemble.

\bigskip
\section*{General \ Setup}

One ensemble corresponding to massless, non interacting \ bosons
and fermions, with equal degrees of freedom, is the massless
$N=1$, $d=4$, Wess-Zumino model. The massless, on shell Wess-Zumino Lagrangian is,%
\begin{equation}
\mathcal{L}=\partial _{\mu }\varphi ^{+}\partial ^{\mu }\varphi +i\overline{\Psi }%
\gamma ^{\mu }\partial _{\mu }\Psi ,  \label{lagra1}
\end{equation}%
with $\Psi$ a Majorana spinor. Writing (\ref{lagra1}) in terms of
two real fields $\phi _{1}$, $\phi _{2}$, $\varphi =\phi _{1}+i\phi
_{2}$, we
obtain,%
\begin{equation}
\mathcal{L}=\partial _{\mu }\varphi _{1}\partial ^{\mu }\varphi
_{1}+\partial _{\mu }\varphi _{2}\partial ^{\mu }\varphi
_{2}+i\overline{\Psi }\gamma ^{\mu }\partial _{\mu }\Psi .
\label{lagracomp}
\end{equation}%
The total space-"time" is of the form $\mathcal{T}\otimes \frac{R^{1}}{Z_{\infty }%
}\times R^{2}$, $\mathcal{T}$ being the "temperature time dimension". Note that $%
Z_{\infty }$ must be a symmetry of the Lagrangian. The compactification of $%
R^{1}$ to $\frac{R^{1}}{Z_{\infty }}$,\ allows us to use generic
boundary conditions for bosons and fermions in the compact
dimension (of course there is some complication with supersymmetry
and boundary conditions, for which, the reader is refered to the
end of this section. The
formulation adopted here is equivalent \cite{Dowker}). These are,%
\begin{eqnarray}
\varphi _{i}(x_{2},x_{3},\tau ,x_{1}) &=&e^{i\pi n_{1}\alpha }\varphi
_{i}(x_{2},x_{3},\tau ,x_{1}+L) \\
~\Psi (x_{2},x_{3},\tau ,x_{1}) &=&e^{i\pi n_{1}\delta }\Psi
(x_{2},x_{3},\tau ,x_{1}+L),  \notag
\end{eqnarray}%
with, $0<\alpha ,\delta <1$, $i=1,2$, $n_{1}=1,2,3...$, while the
finite
temperature boundary conditions are (consistent with KMS relations),%
\begin{eqnarray}
\varphi _{i}(x_{2},x_{3},\tau ,x_{1}) &=&\varphi _{i}(x_{2},x_{3},\tau
+\beta ,x_{1}) \\
~\Psi (x_{2},x_{3},\tau ,x_{1}) &=&-\Psi (x_{2},x_{3},\tau+\beta
,x_{1}),  \notag
\end{eqnarray}%
with $\beta=\frac{1}{T}$. One can check that (\ref{lagracomp}) is
invariant for each representation of
$Z_{\infty }$. The free energy at finite volume for the above ensemble is:%
\begin{equation}
F=F_{bosonic}+F_{fermionic},  \label{eff1}
\end{equation}%
with $F_{bosonic}$, the total bosonic free energy (for $\phi _{1}$,
$\phi _{2}$).
Writing explicitly the free energies we obtain,%
\begin{eqnarray}
F &=&T\sum_{n,m=-\infty }^{\infty } \int \frac{dk^{2}}{(2\pi
)^{2}}\ln
[k^{2}+(2\pi nT)^{2}+(\frac{\pi (2m+\alpha )}{L})^{2}]  \label{eff2} \\
&&-T\sum_{n,m=-\infty }^{\infty }\int \frac{dk^{2}}{(2\pi )^{2}}\ln
[k^{2}+((2n+1)\pi T)^{2}+(\frac{\pi (2m+\delta )}{L})^{2}].  \notag
\end{eqnarray}
For some choices of $\alpha $, and $\delta $, infrared
singularities will occur, and of course an ultraviolet
singularity. The UV will be regularized, using zeta regularization
techniques \cite{elizalde} while, where necessary, infrared
cutoffs will be used to deal with the infrared divergences
\cite{wotzasek}.

\bigskip

\section*{Case $\protect\alpha =1$, $\protect\delta =0$}

If one chooses $\alpha =1$, $\delta =0$, $n_{1}=1$, gets,%
\begin{eqnarray}
F &=&T\sum_{n,m=-\infty }^{\infty } \int \frac{dk^{2}}{(2\pi
)^{2}}\ln
[k^{2}+(2\pi nT)^{2}+(\frac{\pi (2m+1)}{L})^{2}]  \label{eff3} \\
&&-T\sum_{n,m=-\infty }^{\infty } \int \frac{dk^{2}}{(2\pi )^{2}}\ln
[k^{2}+((2n+1)\pi T)^{2}+(\frac{2\pi m}{L})^{2}].  \notag
\end{eqnarray}%
Relation (\ref{eff3}) in $d$ dimensions (at the end $d=4$) is
written,
\begin{eqnarray}
F &=&T\sum_{n,m=-\infty }^{\infty } \int \frac{dk^{d-2}}{(2\pi
)^{d-2}}\ln
[k^{2}+(2\pi nT)^{2}+(\frac{\pi (2m+1)}{L})^{2}]  \label{d} \\
&&-T\sum_{n,m=-\infty }^{\infty } \int \frac{dk^{d-2}}{(2\pi
)^{d-2}}\ln [k^{2}+((2n+1)\pi T)^{2}+(\frac{2\pi m}{L})^{2}], \notag
\end{eqnarray}%
and upon using,%
\begin{equation}
\int \frac{dk^{d}}{(2\pi )^{d}}\ln (k^{2}+a^{2})=-\frac{\Gamma (-\frac{d}{2})%
}{(4\pi )^{\frac{d}{2}}}a^{d},  \label{dimreg}
\end{equation}%
(\ref{d}) becomes,%
\begin{eqnarray*}
F &=&-T\frac{\Gamma (\frac{2-d}{2})}{(4\pi )^{\frac{d-2}{2}}}~%
{\bigg (}\sum_{n,m=-\infty }^{\infty } {\Big (}(2\pi nT)^{2}+(\frac{\pi (2m+1)}{L})^{2}{\Big )}^{%
\frac{d-2}{2}} \\
&&-\sum_{n,m=-\infty }^{\infty } ((2n+1)\pi T)^{2}+(\frac{2\pi m}{L}%
)^{2})^{^{\frac{d-2}{2}}}{\bigg )}.
\end{eqnarray*}%
By introducing the dimensionless parameter $\xi =LT$, we obtain,%
\begin{eqnarray}
F &=&-T\frac{\Gamma (\frac{2-d}{2})}{(4\pi )^{\frac{d-2}{2}}}(\frac{2\pi }{L}%
)^{d-2}{\bigg (}\sum_{n,m=-\infty }^{\infty } {\Big (}\xi ^{2}n^{2}+(m+\frac{1}{2})^{2}{\Big )}^{%
\frac{d-2}{2}}  \label{befepstein} \\
&&-\sum_{n,m=-\infty }^{\infty } {\Big (}\xi ^{2}(n+\frac{1}{2})^{2}+m^{2}{\Big )}^{^{\frac{%
d-2}{2}}}{\bigg )},  \notag
\end{eqnarray}%
In (\ref{befepstein}), the troublesome gamma function $\Gamma
(\frac{2-d}{2})$ appears, which will be canceled later on. Now, with
the aid of the two dimensional form of the Epstein zeta function,
\begin{equation*}
Z_{2}\left\vert
\begin{array}{cc}
g_{1} & g_{2} \\
h_{1} & h_{2}%
\end{array}%
\right\vert (a,a_{1},a_{2})=\sum\limits_{n,m=-\infty }^{\infty}
(a_{1}(n+g_{1})^{2}+a_{2}(m+g_{2})^{2})^{-a}\times \exp [2\pi
i(nh_{1}+mh_{2})],
\end{equation*}%
(\ref{befepstein}) becomes,%
\begin{eqnarray}
F &=&-T\frac{\Gamma (\frac{2-d}{2})}{(4\pi )^{\frac{d-2}{2}}}(\frac{2\pi }{L}%
)^{d-2}{\bigg (}Z_{2}\left\vert
\begin{array}{cc}
0 & \frac{1}{2} \\
0 & 0%
\end{array}%
\right\vert (\frac{2-d}{2},\xi ^{2},1)  \label{afepstein1} \\
&&-Z_{2}\left\vert
\begin{array}{cc}
\frac{1}{2} & 0 \\
0 & 0%
\end{array}%
\right\vert (\frac{2-d}{2},\xi ^{2},1){\bigg )}.  \notag
\end{eqnarray}%
To extend analytically the two dimensional Epstein zeta, to values $Rea<1$, we use the functional equation,%
\begin{align}
&Z_{2}\left\vert
\begin{array}{cc}
g_{1} & g_{2} \\
h_{1} & h_{2}%
\end{array}%
\right\vert (a,a_{1},a_{2}) =  \label{epstein2} \\
&(a_{1}a_{2})^{-\frac{1}{2}}\pi ^{2a-1}\frac{\Gamma (1-a)}{\Gamma (a)}%
\times \exp [-2\pi i(g_{1}h_{1}+g_{2}h_{2})]  \notag \\
&\times Z_{2}\left\vert
\begin{array}{cc}
h_{1} & h_{2} \\
-g_{1} & -g_{2}%
\end{array}%
\right\vert (1-a,\frac{1}{a_{1}},\frac{1}{a_{2}}),  \notag
\end{align}%
and (\ref{afepstein1}) reads:%
\begin{eqnarray}
F &=&-T\frac{\Gamma (\frac{2-d}{2})}{(4\pi )^{\frac{d-2}{2}}}(\frac{2\pi }{L}%
)^{d-2}  \label{afepstein2} \\
&&\times {\bigg (}(\xi ^{2})^{-\frac{1}{2}}\pi ^{\frac{2-2d}{2}}\frac{\Gamma (\frac{d%
}{2})}{\Gamma (\frac{2-d}{2})}Z_{2}\left\vert
\begin{array}{cc}
0 & 0 \\
0 & -\frac{1}{2}%
\end{array}%
\right\vert (\frac{d}{2},\frac{1}{\xi ^{2}},1)  \notag \\
&&-(\xi ^{2})^{-\frac{1}{2}}\pi ^{1-d}\frac{\Gamma (\frac{d}{2})}{\Gamma (%
\frac{2-d}{2})}Z_{2}\left\vert
\begin{array}{cc}
0 & 0 \\
-\frac{1}{2} & 0%
\end{array}%
\right\vert (\frac{d}{2},\frac{1}{\xi ^{2}},1){\bigg )}.  \notag
\end{eqnarray}%
Finally after some algebra we get:%
\begin{eqnarray}
F &=&-\frac{\Gamma (\frac{d}{2})\xi ^{d}\pi ^{-\frac{d}{2}}}{L^{d-1}}
\label{afepstein3} \\
&&\times {\bigg (}Z_{2}\left\vert
\begin{array}{cc}
0 & 0 \\
0 & -\frac{1}{2}%
\end{array}%
\right\vert (\frac{d}{2},1,\xi ^{2})-Z_{2}\left\vert
\begin{array}{cc}
0 & 0 \\
-\frac{1}{2} & 0%
\end{array}%
\right\vert (\frac{d}{2},1,\xi ^{2}){\bigg )},  \notag
\end{eqnarray}%
where the troublesome $\Gamma (\frac{2-d}{2})$ (in $4$ dimensions)
gamma function has been canceled. By introducing the function
$f(\xi )$,
\begin{eqnarray}
f(\xi ) &=&-\frac{L^{d-1}}{\Gamma (\frac{d}{2})\xi ^{d}\pi ^{-\frac{d}{2}}}F
\label{antinvariant} \\
&=&\xi ^{d}{\bigg (}Z_{2}\left\vert
\begin{array}{cc}
0 & 0 \\
0 & -\frac{1}{2}%
\end{array}%
\right\vert (\frac{d}{2},1,\xi ^{2})-Z_{2}\left\vert
\begin{array}{cc}
0 & 0 \\
-\frac{1}{2} & 0%
\end{array}%
\right\vert (\frac{d}{2},1,\xi ^{2}){\bigg )},  \notag
\end{eqnarray}%
we can see that:%
\begin{eqnarray}
f(\frac{1}{\xi }) &=&\frac{1}{\xi ^{d}} {\bigg (}Z_{2}\left\vert
\begin{array}{cc}
0 & 0 \\
0 & -\frac{1}{2}%
\end{array}%
\right\vert (\frac{d}{2},1,\frac{1}{\xi ^{2}})-Z_{2}\left\vert
\begin{array}{cc}
0 & 0 \\
-\frac{1}{2} & 0%
\end{array}%
\right\vert (\frac{d}{2},1,\frac{1}{\xi ^{2}}){\bigg )},
\end{eqnarray}%
or equivalently,
\begin{equation}
f(\frac{1}{\xi })=  Z_{2}\left\vert
\begin{array}{cc}
0 & 0 \\
0 & -\frac{1}{2}%
\end{array}%
\right\vert (\frac{d}{2},\xi ^{2},1)-Z_{2}\left\vert
\begin{array}{cc}
0 & 0 \\
-\frac{1}{2} & 0%
\end{array}%
\right\vert (\frac{d}{2},\xi ^{2},1).  \label{inverse}
\end{equation}
From the last expression we easily obtain:%
\begin{align}
&Z_{2}\left\vert
\begin{array}{cc}
0 & 0 \\
0 & -\frac{1}{2}%
\end{array}%
\right\vert (\frac{d}{2},\xi ^{2},1)-Z_{2}\left\vert
\begin{array}{cc}
0 & 0 \\
-\frac{1}{2} & 0%
\end{array}%
\right\vert (\frac{d}{2},\xi ^{2},1) =  \label{inverse2} \\
&-{\bigg (}Z_{2}\left\vert
\begin{array}{cc}
0 & 0 \\
0 & -\frac{1}{2}%
\end{array}%
\right\vert (\frac{d}{2},1,\xi ^{2})-Z_{2}\left\vert
\begin{array}{cc}
0 & 0 \\
-\frac{1}{2} & 0%
\end{array}%
\right\vert (\frac{d}{2},1,\xi ^{2}){\bigg )}.  \notag
\end{align}%
Thus, combining (\ref{antinvariant}), (\ref{inverse}), (\ref{inverse2}), we get:%
\begin{equation}
f(\xi )=-\xi ^{d}f(\frac{1}{\xi })  \label{antiduality}
\end{equation}%
Relation (\ref{antiduality}) is a realization of an antisymmetry
that the initial ensemble possesses, under the transformation
$L\rightarrow \frac{1}{T}$, or equivalently, $\xi \rightarrow
\frac{1}{\xi }$ (if we can say, this is a kind of anti-duality).
The function $f(\xi )$ is related to the free energy of the system
and, through this, to other thermodynamic quantities of the
ensemble (it is also related to the effective potential).
 Let us check the physical implications of (\ref{antiduality}). The high
temperature free energy is:%
\begin{equation}
F=-T^{d}\pi ^{\frac{2-d}{2}}\Gamma (\frac{d}{2}){\bigg
(}Z_{1}\left\vert
\begin{array}{c}
0 \\
0%
\end{array}%
\right\vert (d)-Z_{1}\left\vert
\begin{array}{c}
0 \\
-\frac{1}{2}%
\end{array}%
\right\vert (d){\bigg )},  \label{hightemp}
\end{equation}
which in $d=4$ is,%
\begin{equation}
F=-T^{4}\pi ^{-2}(\frac{\pi ^{4}}{45}+\frac{7\pi ^{4}}{360}).  \label{4dim}
\end{equation}
The zero temperature Casimir energy for the same ensemble is:%
\begin{equation}
E_{o}=-L^{-d}\pi ^{\frac{2-d}{2}}\Gamma (\frac{d}{2}){\bigg
(}Z_{1}\left\vert
\begin{array}{c}
0 \\
-\frac{1}{2}%
\end{array}%
\right\vert (d)-Z_{1}\left\vert
\begin{array}{c}
0 \\
0%
\end{array}%
\right\vert (d){\bigg )}  \label{casimir}
\end{equation}
which for $d=4$ reads:\bigskip
\begin{equation}
E_{o}=\frac{1}{L^{4}}\pi ^{-2}(\frac{\pi ^{4}}{45}+\frac{7\pi ^{4}}{360})
\label{4dimcasimir}
\end{equation}
Relations (\ref{4dim}) and (\ref{4dimcasimir}) reflect what (\ref%
{antiduality}) expresses, that is the high temperature free energy
of the boson fermion ensemble under consideration, is equal to the
minus zero temperature Casimir energy. This anti-duality found
above was due to choosing antiperiodic boundary
conditions for bosons ($\alpha =1$), and periodic for fermions ($\delta =0$%
), in the compact dimension. There is a mathematical reason for
this choice. Isham \cite{Isham} used these configurations, and
called them twisted fields. These are classified by
$H^{1}(S^{1}{\times R}^{2},Z_{\widetilde{2}})$,
the first Stieffel cohomology group which, in our case is, $H^{1}{(S}^{1}{\times R}^{2}%
{,Z}_{\widetilde{2}}{)=Z}_{2}$ for the spatial section of $\mathcal{T}\otimes \frac{%
R^{1}}{Z_{\infty }}\times R^{2}$. Note that the temperature does not affect
the topological properties of the space \cite{gongcharov} and that $\frac{%
R^{1}}{Z_{\infty }}=S^{1}$. The allowed choices, dictated from $H^{1}{(S}^{1}{%
\times R}^{2}{,Z}_{\widetilde{2}}{)}$, are periodic and antiperiodic
bosons and fermions. Our choice was, antiperiodic bosons and
periodic fermions, which does not influence the supersymmetry
transformations \cite{Isham,
gongcharov}. In addition the Lagrangian is even under $H^{1}{(S}^{1}{\times R}^{2}{%
,Z}_{\widetilde{2}}{)}$ for this choice. In the following section,
antiperiodic fermions and periodic bosons shall be used (similarly
this choice does not influence supersymmetry transformations).

\bigskip
\section*{\protect Case $\protect\alpha =0,~\protect\delta =\frac{1}{2%
}$}

Another case, which will be briefly mentioned, is the choice, $\alpha =0$, $%
\delta =\frac{1}{2}$, $n_{1}=1$ in (\ref{eff2}). Following the steps of the
previous section, one easily obtains the free energy of the system:%
\begin{eqnarray}
F &=&-T\frac{\Gamma (\frac{2-d}{2})}{(4\pi )^{\frac{d-2}{2}}}(\frac{2\pi }{L}%
)^{d-2}{\bigg (}\sum_{n,m=-\infty }^{\infty } {\Big (}\xi
^{2}n^{2}+m^{2}{\Big )}^{\frac{d-2}{2}}
\label{symfree} \\
&&-\sum_{n,m=-\infty }^{\infty } {\Big (}\xi ^{2}(n+\frac{1}{2})^{2}+(m+\frac{1}{2}%
)^{2}{\Big )}^{^{\frac{d-2}{2}}}{\bigg )}.  \notag
\end{eqnarray}
After taking care of the infrared singularity in the bosonic sector \cite{wotzasek}, one obtains finally:%
\begin{eqnarray}
F &=&-\frac{\Gamma (\frac{d}{2})\xi ^{d}\pi ^{-\frac{d}{2}}}{L^{d-1}}
\label{symfreezeta} \\
&&\times {\bigg(}Z_{2}\left\vert
\begin{array}{cc}
0 & 0 \\
0 & 0%
\end{array}%
\right\vert (\frac{d}{2},1,\xi ^{2})-Z_{2}\left\vert
\begin{array}{cc}
0 & 0 \\
-\frac{1}{2} & -\frac{1}{2}%
\end{array}%
\right\vert (\frac{d}{2},1,\xi ^{2}){\bigg )},  \notag
\end{eqnarray}
and using the definition of (\ref{antinvariant}) and the Epstein-zeta
properties we obtain:
\begin{equation}
f(\xi )=\xi ^{d}f(\frac{1}{\xi }).  \label{duality}
\end{equation}
One can easily check that, the high temperature free energy is equal
to the zero temperature Casimir effect, which is expressed from
(\ref{duality}).

\bigskip
\section*{Conclusions}

Studies of the dualities that physical systems have are of great
importance. In this paper, a kind of duality was examined,
expressed in terms of the temperature inversion symmetry of
thermodynamic quantities for a supersymmetric ensemble ($N=1$,
$d=4$ Wess-Zumino with $S^{1}\times R^{2}$ spatial topology, at
finite temperature). This symmetry connects the free energy at
high temperatures (Boltzmann law), with the zero temperature
Casimir energy. Two cases were studied in this work. (a)The case,
with the chiral superfield of the Wess-Zumino model, consisting of
periodic fermions and antiperiodic bosons in the compact
dimension. In this case we found the function $f(\xi )$, defined
in (\ref{antinvariant}), being antisymmetric under the
transformation $\xi \rightarrow \frac{1}{\xi }$, {\it{i.e.}}~
$f(\xi )=-\xi ^{d}f(\frac{1}{\xi })$. Thus the high temperature
free energy of the boson fermion ensemble is equal to the minus
zero temperature Casimir energy. (b)The case in which periodic
bosons and antiperiodic fermions in the compact dimension, were
used. In that case, the function $f(\xi )$ was found symmetric
under the transformation $\xi \rightarrow \frac{1}{\xi }$,
{\it{i.e.}}~ $f(\xi )=\xi ^{d}f(\frac{1}{\xi })$. Consequently,
the high temperature free energy is equal to the zero temperature
Casimir energy. The function $f(\xi )$, is related to the free
energy of the system under consideration and the results show, how
boundary conditions can affect the quantum structure of the
system, in terms of $f(\xi )$. Finally, one can easily observe
that if ${\xi }=1$, then the free energy of the system, composed
of periodic fermions and antiperiodic bosons, becomes equal to
zero. The ${\xi }=1$ limit corresponds to the case
$L=\frac{1}{T}$. If interactions were added, then, maybe, this
limit could connect phase transitions at some critical
temperature, with a dual critical length. If the free energy
remains zero at this limit, the question is what the physics of a
system with zero free energy at a temperature $T\neq 0$, is (one
system having zero free energy is the d5 Ads Black hole with Ricci
flat horizons. The zero of the free energy occurs for the Hawking
temperature $T_{BH}=\frac{1}{ln_{BH}}$ ($l$ the radius of the
asymptotic Ads space), and determines the critical point of the
transition, Ads black hole to Ads soliton \cite{odintsov}).

\bigskip

\bigskip

\section*{Acknowledgements}

\noindent This research was co-funded by the European Union in the framework
of the Program PYTHAGORAS-I of the "Operational Program for Education and
Initial Vocational Training" (EPEAEK) of the 3rd Community Support Framework
of the Hellenic Ministry of Education, funded by 25{\%} from national
sources and by 75{\%} from the European Social Fund (ESF).

\end{document}